\documentclass{iopart}

\usepackage{graphicx}
\usepackage{textcomp}
\usepackage{gensymb}
\usepackage[square,numbers,sort&compress]{natbib}

\newcommand{\dd}{\ensuremath{\mathrm{d}}}

\newcommand{\diff}[2]{\ensuremath{\frac{\dd {#1}}{\dd {#2}}}}

\newcommand{\mean}[1]{\ensuremath{\left\langle {#1}\right\rangle}}

\newcommand{\um}[1]{#1~\textmu m}
\newcommand{\us}[1]{#1~\textmu s}
\newcommand{\eqref}[1]{\eref{#1}}

\newcommand{\keff}{\ensuremath{k_\mathrm{eff}}}
\newcommand{\meff}{\ensuremath{m_\mathrm{eff}}}

\usepackage{etoolbox}
\patchcmd{\numparts}{\addtocounter{equation}{1}}{\refstepcounter{equation}}{}{}

\def\3He{$^3$He}
\def\4He{$^4$He}

\begin{document} 

\title{Electromechanical feedback control of nanoscale superflow}

\author{E. Varga, J. P. Davis}
\address{Department of Physics, University of Alberta, Edmonton, Alberta T6G 2E1, Canada}
\ead{ev@ualberta.ca, jdavis@ualberta.ca}

\begin{abstract}
  Superfluid \4He is a promising material for optomechanical and electromechanical applications due to its low acoustic loss. Some of the more intriguing aspects of superfluidity -- the macroscopic coherence, topological nature of vorticity, and capability of supporting non-classical flows -- remain, however, poorly explored resources in opto- and electro-mechanical systems. Here, we present an electromechanical coupling to pure superflow inside a nanofluidic Helmholtz resonator with viscously clamped normal fluid. The system is capable of simultaneous measurement of displacement and velocity of the Helmholtz mechanical mode weakly driven by incoherent environmental noise. Additionally, we implement feedback capable of inducing self-oscillation of the non-classical acoustic mode, damping the motion below the ambient level, and tuning of the mode frequency.
\end{abstract}


\section{Introduction}
\label{sec:intro}

Superfluid helium (He-II), the phase of liquid helium that exists below approximately 2.2 K, has several properties that make it an appealing material for implementing mechanical elements in optomechanical and electromechanical systems. Cryogenic compatibility, low acoustic loss, and the ability to fill arbitrary volumes and wet arbitrary surfaces have already found multiple applications, e.g., in prototype tabletop detectors of gravitational waves \cite{Lorenzo2014,Singh2017,Vadakkumbatt2021}; as the acoustic medium in a fibre-based optical cavity demonstrating sideband asymmetry \cite{Kashkanova2016,Kashkanova2017,Shkarin2019}; proposed levitated helium drop experiments \cite{Childress2017}; for flow-mediated actuation of micromechanical resonators \cite{McAuslan2016}; or demonstrating coupling to third-sound surface waves on thin helium films \cite{Harris2016,Sachkou2019}, including strong coupling and mechanical lasing \cite{He2020}.

He-II is an intriguing mechanical element for reasons beyond its low dissipation and ease of use, however. At temperatures above approximately 1 K, He-II behaves as a mixture of two components \cite{Tilley_book} -- normal and superfluid -- each with its own velocity field. In addition, vorticity in the superfluid component can exist only in the form of angstrom-thick topological defects (quantized vortices). These effects lead to rich dynamical behaviour not present in typical mechanical elements. Non-classical flows with non-zero relative velocities between the two fluid components occur, where the complex dynamics of the tangle of quantized vortices is a subject of intense study \cite{Barenghi2014c,Skrbek2021}. Confined to two dimensions, formation of large-scale Onsager vortex clusters was observed in a freely-evolving point-vortex system \cite{Sachkou2019} and bi-stability of the macroscopic turbulent state has been observed in a driven system \cite{Varga2020}, related to large-scale polarisation of the vortex system. Macroscopic coherence also allows implementation of superfluid Josephson junctions \cite{Avenel1985}, superfluid quantum interference devices (SHeQUIDs) \cite{Sato2011}, and possibly mechanical qubits \cite{Sfendla2021}.

In this work, we present an implementation of an electromechanical system coupled to \emph{pure superflow} of He-II confined to 450 nm in a nanofluidic Helmholtz resonator \cite{Varga2020,Shook2020,Souris2017}, where the normal fluid is rendered immobile by viscous clamping (a 4th-sound acoustic mode \cite{Tilley_book}). We detect the helium flow driven by ambient environmental noise (i.e., the Helmholtz resonator acts as a superfluid microphone). The detection principle of our system relies on mixing the mechanical fluctuations with a monochromatic carrier wave resulting in mechanical sidebands in a manner reminiscent of cavity-less optomechanics \cite{Renninger2018}, albeit operating at much lower frequencies of the elelctromagnetic spectrum.

Since the flow of helium in our nanofluidic resonator can be electrostatically driven, it is possible to use the measured information about the mechanical motion to dampen or amplify the motion. Feedback control of mechanical motion is an important tool for noise reduction in optomechanical sensors \cite{Tsang2010}, frequency tuning to the range of optimal sensitivity \cite{Whittle2021}, or cooling of the thermal occupation of the mechanical motion (e.g., \cite{DUrso2003,Poggio2007,Kim2017}), -- even achieving near-ground-state phonon occupation of a macroscopic 10~kg object \cite{Whittle2021}.

In our case, using feedback control we show reduction of the root-mean-square displacement of the helium flow by an order of magnitude compared to the ambient value, amplification to the point of coherent self-oscillation, and frequency tuning. The electromechanical control and readout of the nanofluidic resonator presented here provides a stepping stone towards superfluid electromechanics with engineered nanostructures capable of exploiting a wider breadth of superfluid phenomena in nanoscale superflows.

\section{Helmholtz Mechanical Mode and the Experimental Setup}
\label{sec:experiment}

\begin{figure}
    \centering
    \includegraphics{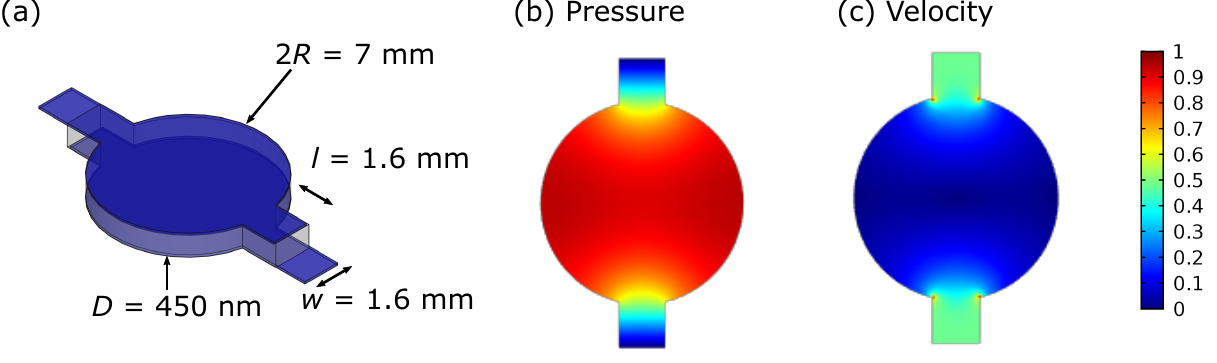}
    \caption{The nanofluidic Helmholtz resonator. (a) Sketch of the device used in the study. Lateral dimensions are to scale, thickness is exaggerated. The central circular basin is connected to the outside bath of He-II through the two side channels. The helium is uniformly confined throughout the device apart from a \um{10} gap around the edges of the 50~nm thick Al electrodes (dark blue). (b) The pressure oscillation in the mode. (c) The flow velocity in the mode (color scale in arbitrary units).}
    \label{fig:helmholtz_resonator}
\end{figure}

The studied flow of He-II occurs in the Helmholtz resonator, which consists of two identical channels connecting the helium bath, in which the resonator is immersed, to the central circular basin as shown in Fig.~\ref{fig:helmholtz_resonator}. The resonator is similar to resonators used in previous studies \cite{Varga2020,Shook2020}. Contrary to previous studies, however, we detect incoherent motion driven by ambient noise rather than coherently driving the mode and in addition we apply feedback to control this motion.

The helium in the device is confined uniformly (apart from a \um{10} gap around the side walls, which corresponds to about 0.6\% of the total area of the resonator) between the two aluminum electrodes sputter-deposited on top and bottom walls of the confined volume. The confining volume is etched into single-crystal quartz to an overall confinement of $D\approx 450$ nm (see \cite{Souris2017} for details on the fabrication). The mode can be described as a mass-on-a-spring oscillator \cite{Varga2020} with effective mass $\meff = 2wlD\rho_s$ with channel width and length $w$ and $l$, respectively (see Fig.~\ref{fig:helmholtz_resonator}) and effective spring constant $\keff$, resulting in mode frequency
\begin{equation}
    \label{eq:frequency}
    \Omega_{m0} = \sqrt{\frac{\keff}{\meff}} = \sqrt{\frac{2wD}{l\rho}\frac{\rho_s}{\rho}\frac{k_p}{4A^2(1 + \Sigma)}},
\end{equation}
where $k_p$ is the effective stiffness of the mean deflection of the quartz walls of the basin, $A=\pi R^2$ is the area of the basin, and $\Sigma = \chi_B D k_p/(4A)$, where $\chi_B$ is the bulk compressibility of helium. The $\rho_s/\rho$-dependence of the mode frequency has been exploited for the identification of new phases of \3He \cite{Shook2020} and allows for easy tunability of the mechanical frequency by adjusting the temperature or pressure of the helium bath.

The detection of the flow of helium in the device is facilitated by the parallel plate capacitor, formed by the Al electrodes deposited on the top and bottom walls of the confined volume, of capacitance $C_0\approx 1$~nF. As the helium moves in and out through the channels, the pressure in the basin oscillates, which in turn modulates the capacitance of the device. The fluctuating part of the capacitance $\delta C$ is directly proportional to the displacement of the liquid in the channels connecting the basin of the resonator to the external bath \cite{Varga2020} as
\begin{equation}
\label{eq:Cy-coupling}
    \delta C = g_C C_0 y,
\end{equation}
where $y$ is the helium displacement in the channels with sensitivity 
\begin{equation}
    \label{eq:gc}
    g_C = \frac{2w\rho_s}{A\rho}\left(1 - 2\frac{\varepsilon - 1}{\varepsilon}\Sigma\right)\frac{1}{1 + 2\Sigma},
\end{equation}
where $\varepsilon$ is the dielectric constant of $^4$He.

\begin{figure*}
    \centering
    \includegraphics[width=\linewidth, keepaspectratio]{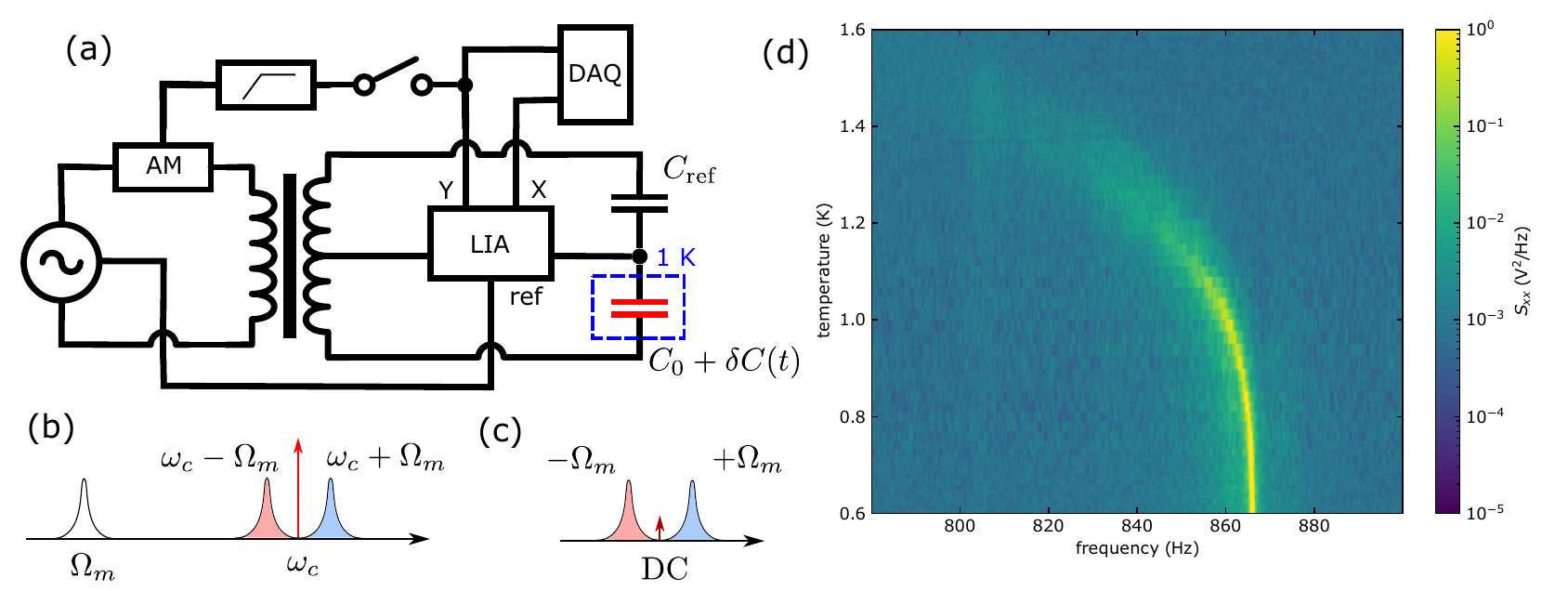}
    \caption{The experimental setup. (a) Schematic drawing of the detection circuit and the feedback loop. The mechanical motion causes the capacitance of the Helmholtz resonator to fluctuate, which induces sidebands around the carrier wave. The current signal is then demodulated by the lock-in amplifier (LIA), referenced to the frequency of the carrier, and the noise on the two quadratures is then digitized and optionally fed into amplitude modulation of the carrier. (b) The spectrum of the mechanical mode at frequency $\Omega_m$ and the spectrum of the current passing through the device (shaded). The mechanical motion modulates the carrier signal at frequency $\omega_c$ and creates two sidebands at $\omega_c \pm \Omega_m$. (c) The spectrum of the current demodulated by the LIA referenced to the carrier frequency. The power at the carrier frequency is strongly reduced by tuning the reference capacitor $C_\mathrm{ref}$. The fluctuations at the output of the LIA contain the information on the displacement and velocity of helium flow in the nanofluidic Helmholtz mode. (d) Power spectral density of the detected noise as a function of temperature and frequency. The noise-driven Helmholtz mode can be clearly seen below approximately 1.5 K.}
    \label{fig:system}
\end{figure*}

We observe flow within the Helmholtz resonator through the detection of sidebands induced on a high-frequency carrier signal (high relative to the mechanical frequency), as shown in Fig.~\ref{fig:system}(a-c). The Helmholtz resonator acts as a capacitor with time-varying capacitance $C(t) = C_0 + \delta C(t)$; applying a voltage $u_c(t) = U_c\cos(\omega_c t)$ across its electrodes results in current given by $I(t) = \dd (C(t)u(t))/\dd t$.

By wiring the resonator in a bridge circuit (Fig.~\ref{fig:system}(a)) balanced by a reference capacitor $C_\mathrm{ref}$, the component of the signal due to static capacitance of the device can be strongly reduced and the resulting current through the detector is
\begin{equation}
    \label{eq:current}
    I(t) = -\omega_c U_c\delta C(t)\sin(\omega_c t) + U_c\delta\dot{C}(t)\cos(\omega_c t).
\end{equation}
Due to the tuned balance of the bridge circuit, only the relatively weak signal of the mechanical sidebands enters the amplification and detection chain, which would be otherwise overloaded by the carrier signal (we note that a similar detection scheme was used in the past for the investigation of surface waves on \3He and \4He \cite{Manninen2014,Manninen2014a}).

The current signal is demodulated using a lock-in amplifier with the time constant set sufficiently short ensuring that the mechanical sidebands are within the bandwidth of the low-pass filter. The two outputs of the lock-in amplifier are $X_\phi(t) = G_{IV}\gamma_L\{I(t)\cos(\omega_c t + \phi)\}$ and $Y_\phi(t) = G_{IV}\gamma_L\{I(t)\sin(\omega_c t + \phi)\}$ or, in terms of displacement $y$ and velocity $\dot y$ (dropping the terms at frequencies $2\omega_c$) of the helium flow using \eqref{eq:Cy-coupling},
\begin{numparts}
\label{eq:XY}
\begin{eqnarray}
\label{eq:X}
    X_\phi(t) = \frac{1}{2}G_{IV}C_0g_CU_c\gamma_L\{\dot y(t) \cos\phi + \omega_c y\sin\phi\}\\
\label{eq:Y}
    Y_\phi(t) = \frac{1}{2}G_{IV}C_0g_CU_c\gamma_L\{\dot y(t) \sin\phi - \omega_c y\cos\phi\}.
\end{eqnarray}
\end{numparts}Here $G_{IV} = 5\times 10^5$ V/A is the total gain of the current preamplifier and the lock-in amplifier, $\phi$ is the adjustable phase of the lock-in, and $\gamma_L\{\cdot\}$ stands for a low-pass filter with complex frequency response $\gamma_L(\omega) = 1/(1 + i\omega \tau_L)$ (in our case $\tau_L=$\us{100} corresponding to -6 dB cutoff frequency of about 1600 Hz). The frequency of the carrier signal (in our case $\omega_c/2\pi = 31$ kHz) is arbitrary, as no resonant enhancement takes place, and is chosen to lie in a region relatively free of noise and within the bandwidth of the amplifiers used. Note that for $\phi=0$ we have $X_0(t) \propto y(t)$ and $Y_0(t)\propto \dot y(t)$, which in principle allows simultaneous detection of displacement and velocity of the helium flow. Continuous velocity measurement in optomechanical systems is typically significantly more challenging \cite{Vyatchanin2016}. It is interesting to note that the relative sensitivity of the detection of the velocity and the displacement depends on the carrier frequency, i.e. for $\omega_c \to 0$ only the flow velocity would be detected and for $\omega_c \to \infty$ the displacement dominates the signal.

However, the case just presented applies to an idealized circuit with a purely capacitive load and negligible phase rotation of the carrier signal due to filters and amplifiers. Any resistive elements, combined with an additional phase rotation due to the detection electronics, will significantly complicate the expression for the detected current \eqref{eq:current} and the the information on the displacement and its time derivative no longer decomposes neatly into the two quadratures of the carrier signal. These parasitic effects are dependent on the experimental parameters and are difficult to characterize sufficiently accurately to be fully compensated in the present experiment. However, in terms of the power spectral densities $S_{zz}(\omega) = \left|\tilde z(\omega)\right|^2$ (where $\tilde{}$ denotes the Fourier transform) it can be shown (see Appendix) that, to the first order the combined power spectral density is independent of the parasitic resistance $R$, i.e.,
\begin{equation}
    \label{eq:Sxxyy}
    S_{XX}(\omega) + S_{YY}(\omega) = \frac{1}{4}G_{IV}^2U_c^2(\omega_c^2+\omega^2)C_0^2g_C^2\left|\gamma_L(\omega)\right|^2S_{yy}(\omega),
\end{equation}
which is also independent of possible spurious rotation of the carrier phase and contains only well-characterized known quantities. 

The mechanical mode can be driven directly by applying an AC voltage $u$ across the electrodes of the device. However, since the electrostatic force $F_\mathrm{es}\propto u^2$, this requires either reducing the frequency of the drive signal to half the frequency of the mode or applying additional DC bias. Changing the frequency of the drive complicates the implementation of the feedback loop and introduction of bias complicates the measurement circuit, therefore we adopt an alternative method of forcing by amplitude-modulating the carrier wave, which does not suffer from the above mentioned issues.

Amplitude modulation of the carrier voltage takes the form $u_c^\mathrm{AM}(t) = \frac{1}{2}U_c[1 + da(t)]\cos(\omega_c t)$, where $d$ is the modulation depth (ranging from 0 to 100\%) and $a(t)$ is the modulating signal, $|a(t)| \leq 1$. The electrostatic force across the plates of the capacitor is given by $F_\mathrm{es} = C_0(u_c^\mathrm{AM})^2/2D$. Neglecting the components of the force far off mechanical resonance (DC and at multiples of $\omega_c$), the force reduces to $F_\mathrm{es} = C_0U_c^2da(t)/4D$.

Finally, the feedback loop is implemented by setting $a(t) = G_\mathrm{AM}\gamma_H\{Y_\phi(t)\}$, where $Y_\phi$ is one of the lock-in outputs (with lock-in phase at $\phi$), and $G_\mathrm{AM} = 0.2$ V$^{-1}$ is a fixed gain of the input stage of the amplitude modulator, and $\gamma_H$ stands for a RC high-pass filter with complex frequency response $\gamma_H(\omega) = i\omega\tau_H/(1 + i\omega\tau_H)$ with the time constant $\tau_H$~=~\us{27}. The total feedback gain $G_\mathrm{FB}$ is directly proportional to the amplitude modulation depth $d$, through which it is adjustable.

\section{Noise-Driven Motion and Feedback}
\label{sec:results}

The temperature dependence of the power-spectral density of the detected noise is shown in Fig.~\ref{fig:system}(d). The prominent signal corresponds to the Helmholtz mechanical mode which follows the temperature dependence of the superfluid fraction $\rho_s/\rho$ through \eqref{eq:frequency}. In this case, the motion is driven only by the environmental noise (with the feedback loop disconnected). The coupling of the mode to environmental noise occurs through homogeneous pressure fluctuations in the bath (caused primarily by the vibrations emanating from a cryocooler attached to the refrigerator), which can drive the Helmholtz mode since the flow through the channels is symmetric with respect to the bath. This relatively significant drive prevented us from observing the equilibrium thermal excitation of the mechanical mode.

Even though the observed motion is not thermomechanical in origin, the random driving force is unrelated to the detection circuit and it is possible to apply feedback to control it. The motion can be damped or amplified by applying a force proportional to the measured velocity, or the frequency of the mode can be modified by applying a force proportional to the measured displacement or acceleration.

The effect of the feedback can be understood directly through the equation of motion for the helium displacement $y$ (in Fourier space),
\begin{equation}
\label{eq:helium-eom}
    -\meff\omega^2 y + \keff y + i\omega\meff\Gamma_0 y = F_\mathrm{noise} + F_\mathrm{FB},
\end{equation}
where $\meff$ is the effective mass of the moving helium, $\keff$ is the effective spring constant, and $\Gamma_0$ is the intrinsic damping of the resonance. The random forcing $F_\mathrm{noise}$ due to environmental noise is assumed to be spectrally flat in the bandwidth of the mechanical resonance (in a noise-free system this would reduce to the forcing due to thermal fluctuations). The force due to feedback is
\begin{equation}
\label{eq:feedback-force}
    F_\mathrm{FB} = y G_\mathrm{FB}\gamma \left(\Omega_C\cos\phi + i\omega\sin\phi\right) = yg(\phi, d),
\end{equation}
where $G_\mathrm{FB}\propto d$ is the overall gain of the feedback loop proportional to the AM depth $d$; $\gamma = \gamma_L\gamma_H = i\omega\tau_H/(1 + i\omega\tau_H)(1 + i\omega\tau_L)$ is the response of the combined low-pass filter of the lock-in amplifier and the high-pass filter between lock-in output and the AM input; and $\phi$ is the lock-in phase (i.e., relative phase between the carrier wave and the lock-in reference signal). The overall gain and phase response of the feedback loop is characterized by the function $g(\phi, d)$, which depends on the lock-in phase  $\phi$ and AM depth $d$ (through $G_\mathrm{FB}\propto d$). The feedback modifies the effective susceptibility of the resonance to $\chi(\omega) = \left[(\Omega_m^2 - \omega^2) + i\Gamma\omega\right]^{-1}$, where
\begin{equation}
\label{eq:feedback-modified-frequency}
    \Omega_m^2 = \Omega_{m0}^2 - \Re g,
\end{equation}
and
\begin{equation}
\label{eq:feedback-modified-gamma}
    \Gamma = \Gamma_0 - \Im\frac{g}{\omega}.
\end{equation}
The power spectral density of the detected displacement noise is then
\begin{equation}
\label{eq:Syy}
    S_{yy}(\omega) = \frac{S_{FF}}{m_\mathrm{eff}^2[(\Omega_m^2 - \omega^2)^2 + \Gamma^2\omega^2]},
\end{equation}
where $S_{FF}$ is the power spectral density of the force noise $F_\mathrm{noise}$.

\begin{figure}
    \centering
    \includegraphics[width=\linewidth, keepaspectratio]{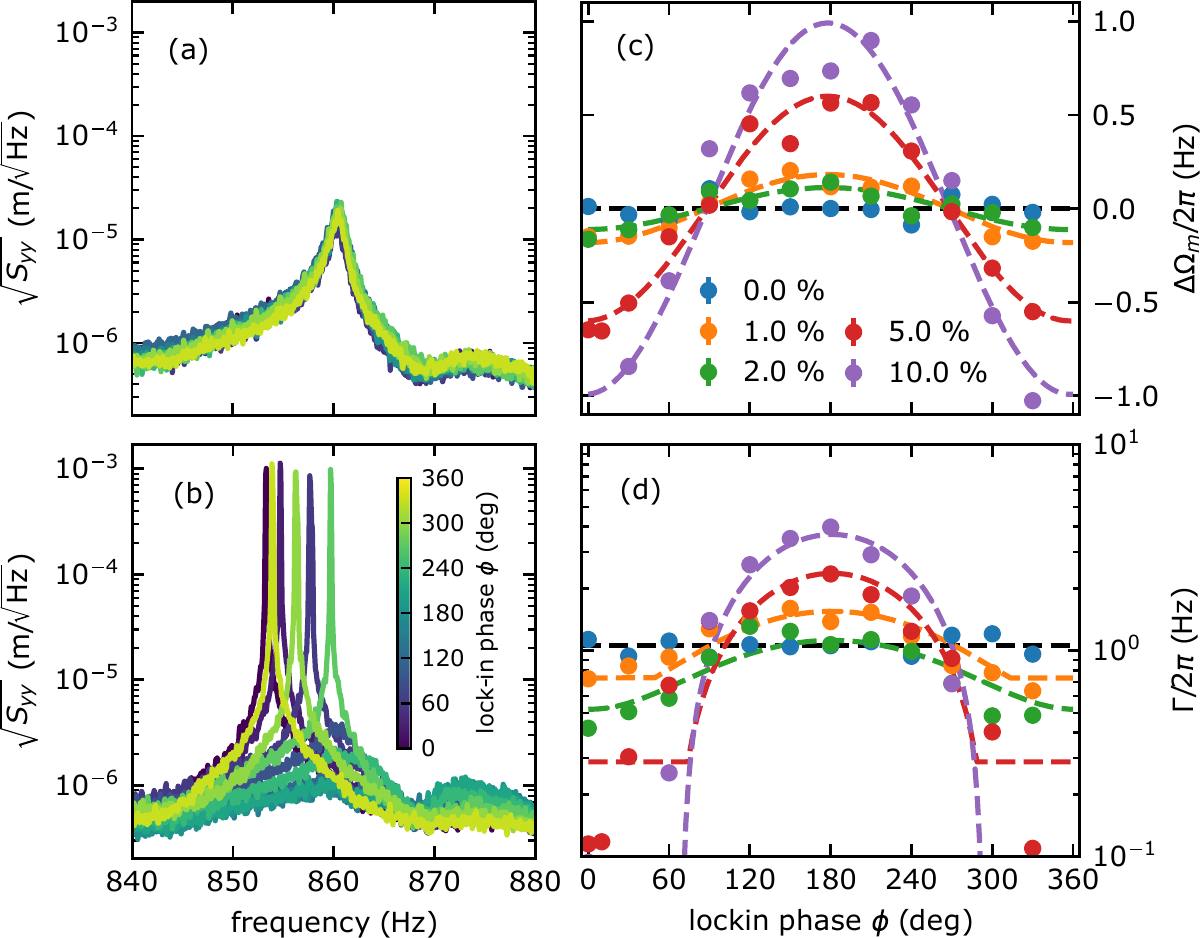}
    \caption{Effect of the detection phase $\phi$ on the observed displacement spectral densities at 1 K without (a) and with (b) feedback (50\% AM depth). In the undriven case the spectral density is independent of the detection phase. With feedback, the flow can be strongly enhanced or suppressed. With increasing feedback gain both the central frequency (c) and linewidth (d) are affected. The dashed lines in (c) and (d) are fits to \eref{eq:feedback-modified-frequency} and \eref{eq:feedback-modified-gamma}, respectively. The colorbar in (b) identifies the detection phase in both (a) and (b) and the legend in (c) indicates the feedback gain in terms of the modulation depth for both (c) and (d).}
    \label{fig:feedback}
\end{figure}

The fits to \eqref{eq:feedback-modified-frequency} and \eqref{eq:feedback-modified-gamma} in Fig.~\ref{fig:feedback}(c,d) show good agreement. For sufficiently high feedback gains, the linewidth \eqref{eq:feedback-modified-gamma} would decrease to negative values. At this threshold the system starts to oscillate coherently at its mechanical frequency (sometimes referred to as mechanical or phonon lasing \cite{Pettit2019}, although we note that in our case it is the external feedback loop that drives the mechanical motion, rather than instability resulting from parametric coupling of the mechanical motion to a pump \cite{Kippenberg2005,Mahboob2013,Potts2021}). The amplitude of the oscillations is limited to a finite value (and the linewidth remains positive) via additional non-linear dissipation, which in our case is most likely due to turbulence that develops in the helium flow \cite{Varga2020}. We note that this is the first example of superfluid 4th sound lasing and, to the best of our knowledge, the first case where the limiting nonlinear dissipation is due to quantum turbulence.

\begin{figure}
    \centering
    \includegraphics[width=\linewidth, keepaspectratio]{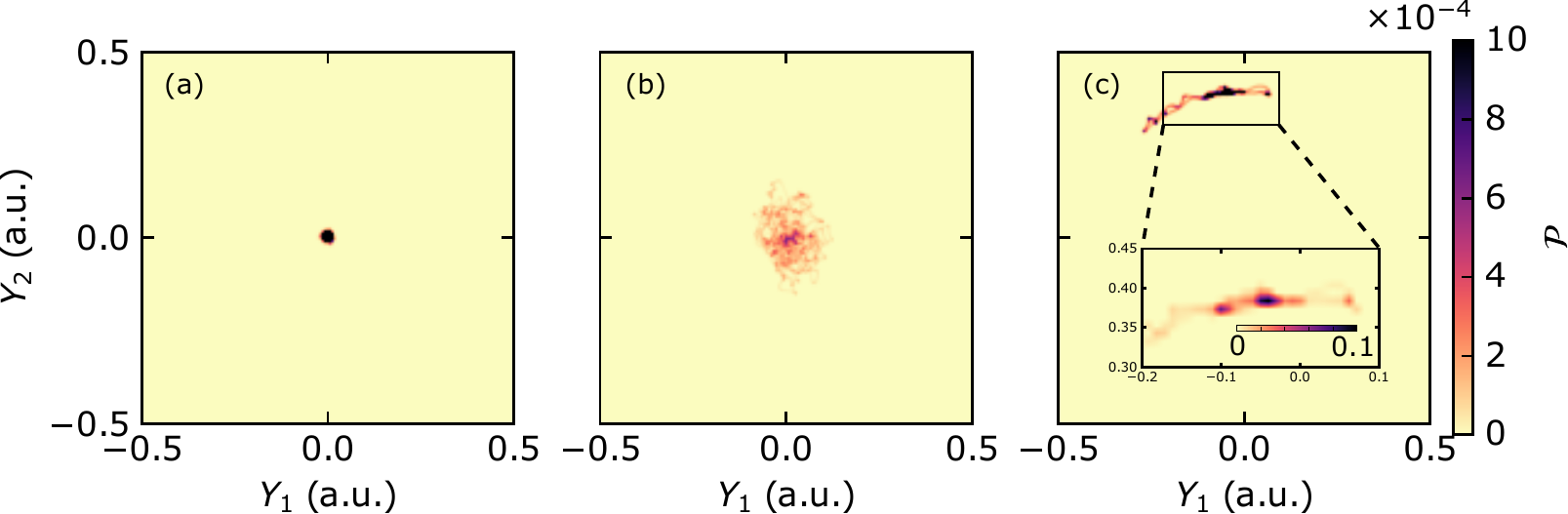}
    \caption{Two dimensional histograms of the two quadratures of the helium displacement given by \eref{eq:y-quadratures} at 0.7 K (i.e., helium displacement time series demodulated near the mechanical frequency). (a) Feedback damping ($\phi=180\degree$, $d=5\%$). (b) Ambient conditions with no feedback. (c) Coherent self-oscillations ($\phi=0\degree$, $d=5\%$). The inset in (c) shows a magnified portion of the histogram with the probability color scale expanded by a factor of 100. With large probability the oscillations remain coherent at a constant amplitude and phase. The phase noise that smears the histogram into an arc is most likely caused by frequency fluctuations due to pressure and temperature instabilities.}
    \label{fig:quadratures}
\end{figure}

The average energy of the helium flow can be either increased or reduced with respect to the noise-driven value, depending on the phase of the lock-in detection, thus increasing or decreasing the root-mean-square (RMS) displacement of the helium. This can be seen as a change in the amplitude of the PSDs in Fig.~\ref{fig:feedback}b (compare with Fig.~\ref{fig:feedback}a which shows no phase-dependence of the PSD with the feedback loop disconnected). More directly, the stochastic flow can be represented via quadratures $Y_1, Y_2$ as 
\begin{equation}
    \label{eq:y-quadratures}
    Y(t) = Y_1(t)\cos(\Omega_m t) + Y_2(t)\sin(\Omega_m t),
\end{equation}
shown in Fig.~\ref{fig:quadratures} for the $Y$-output of the lock-in. The stochastic Brownian-like motion observed without feedback (Fig.~\ref{fig:quadratures}b) can be significantly reduced with feedback damping (Fig.~\ref{fig:quadratures}a) or amplified to the point of coherent oscillations (Fig.~\ref{fig:quadratures}c). Coherent oscillations of constant amplitude and phase appear as a single point away from origin.

\begin{figure}
    \centering
    \includegraphics{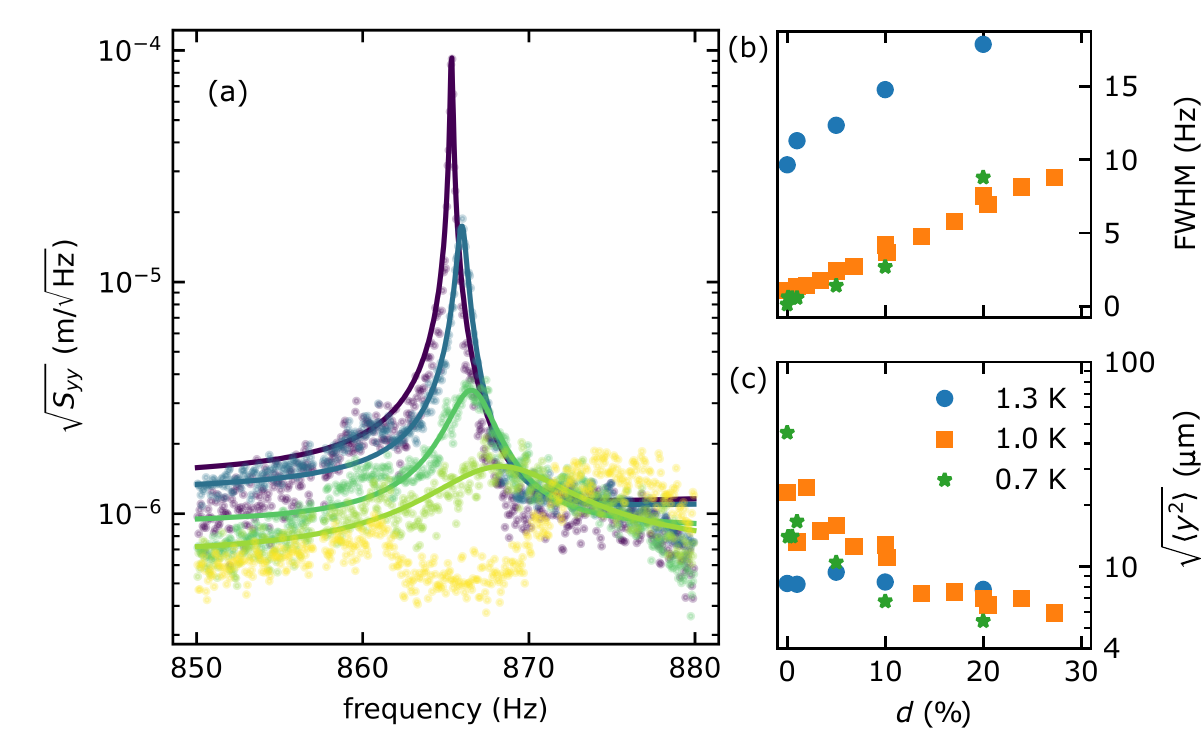}
    \caption{(a) Power spectral densities of the helium flow displacement at 0.7 K for $\phi=180\degree$ and several gains ($d = $ 0, 1, 10, 20 and 50\%, top to bottom). Solid lines are fits to \eqref{eq:Syy}. The lowest dataset with the highest feedback gain (the yellow points without a fit) shows evidence of noise squashing.  (b, c) The linewidth of the resonance and the RMS displacement of the helium flow, respectively, at $\phi=180\degree$ (corresponding to feedback damping) as a function of the feedback gain given by the depth of the amplitude modulation.} 
    \label{fig:rms-displacement}
\end{figure}

Thanks to the relative simplicity of our detection circuit, the electromechanical calibration can be carried out directly based on the known capacitance sensitivity \eqref{eq:gc}, and total gain of the detection instruments (in most electromechanical and optomechanical systems this calibration typically requires invoking the fluctuation-dissipation theorem \cite{Hauer2013} if thermomechanical motion is observable). The calibrated root-mean-square displacement of the helium as a function of the feedback gain is shown in Fig.~\ref{fig:rms-displacement}. The mean-square displacement is calculated as the area under the curve of the PSD \eqref{eq:Syy}
\begin{equation}
\label{eq:rms-y}
    \mean{y^2} = \int_0^\infty S_{yy}(\omega)\dd\omega = \frac{\pi S_{FF}}{2m_\mathrm{eff}^2\Omega_m^2\Gamma},
\end{equation}
where $S_{FF}/m_\mathrm{eff}$, $\Gamma$ and $\Omega_m$ are obtained by fitting the experimental data (note that the slight asymmetry of the peaks is due to interference with the background, which slightly modifies the line shape from \eref{eq:Syy} to \eref{eq:noisy-Syy} derived in the Appendix). Even though the environmental noise keeps the mode relatively strongly driven, the feedback damping reduces the RMS displacement of the helium flow at 0.7 K by approximately an order of magnitude. For high feedback gains, so-called noise squashing \cite{Poggio2007} becomes apparent, where the mechanical motion becomes driven via feedback of the measurement noise with which it destructively interferes. This is manifested as a reduction of the PSD below the local noise level (see the bottom yellow data set in Fig.~\ref{fig:rms-displacement}(a)). At this point the expression \eqref{eq:rms-y} for the rms displacement is no longer accurate as the mechanical motion is strongly driven by the feedback of the measurement noise \cite{Poggio2007}. In our case the rather strong and non-trivial background prevented an accurate fitting, therefore we limit the analysis of the displacement to weaker feedback gains $d \leq 30\%$

Finally, it is interesting to compare the expected height of PSD peak resulting from the thermomechanical motion with our current noise background. Forcing by pure thermal fluctuations follows from the fluctuation-dissipation theorem \cite{Hauer2013} $S_{FF}~=~4k_BT\Gamma_0m_\mathrm{eff}$. For the effective mass of the Helmholtz resonance $m_\mathrm{eff}~\approx~\rho_\mathrm{He}wlD = 167$~ng, frequency $\Omega_{m0}/2\pi = 865$ Hz, and decay rate $\Gamma_0/2\pi \approx 0.1$~Hz, we find at $T = 0.7$~K that $\sqrt{S_{yy}^\mathrm{max}} = S_{FF}/m_\mathrm{eff}^2\Gamma_0^2\Omega_{m0}^2 \approx 10^{-10}$~m/$\sqrt\mathrm{Hz}$, about 4 orders of magnitude below our current noise level. While clearly unobservable in the present experiment, the noise level is not a fundamental limitation of the experimental scheme and can be reduced substantially. The Helmholtz mechanical mode can be designed (using multiple interconnected basins) with no net flow in or out of the confined volume, which will substantially reduce the susceptibility to pressure fluctuations in the surrounding bath. Second, the detection sensitivity of the displacement is proportional to the carrier frequency (see \eqref{eq:XY}), which can be increased from the current 31 kHz. Finally, the cryogenic system used in this study suffered from significant mechanical noise due to a cryocooler attached to the refrigerator. The outlined issues will be addressed in a future experiment.

\section{Conclusions} 
This work presents electromechanical coupling to the 4th-sound acoustic mode of superfluid helium in a nanofluidic Helmholtz resonator. We detect the mode driven by ambient incoherent noise at finite temperatures with the normal fluid component viscously clamped by nanoscale confinement. Feedback control of this non-classical acoustic mode is possible, enabling cold damping, self-oscillation (`mechanical lasing'), and tuning of the mechanical frequency.

Detection of the thermomechanical motion is in principle possible and will be the goal of future experiments.
Due to the presence of dissipative coupling of the superflow to the stationary normal component (mutual friction \cite{Tilley_book}), mediated by the presence of pinned quantized vortices \cite{Awschalom1984a}, this system will offer an interesting possibility of studying the thermal fluctuations in an out-of-equilibrium state stabilized by the topological nature of the quantized vortices. Finally, we note that the relatively low carrier frequencies and high reactive load in the present experiment strongly limit heat input into the system, making it compatible with ultra-low temperature superfluid $^3$He experiments, where nanofluidic Helmholtz resonators have already been used \cite{Shook2020}.

\section{Acknowledgments}
We are grateful to C. A. Potts for fruitful discussions and G. G. Popowich for technical assistance. This work was supported by the University of Alberta; the Natural Sciences and Engineering Research Council, Canada (Grant No.~RGPIN-04523-16); and the Alberta Quantum Major Innovation Fund. 

\section*{Appendix}

A more realistic circuit will include resistance in the unknown capacitance arm in the bridge circuit in Fig.~\ref{fig:system}. We can model this resistance as a discrete resistor $R$ in series with the device capacitance $C(t)$. The charge $q$ on the device will evolve according to a differential equation
\begin{equation}
\label{eq:charge}
    q(t) + RC(t)\diff{q(t)}{t} = U_cC(t)\cos(\omega_c t),
\end{equation}
with the current being $I(t) = \dd q/\dd t$. Assuming that $C(t) = C_0 + \delta C\cos(\Omega_m t)$ we solve \eqref{eq:charge} with an ansatz
\begin{equation}
\label{eq:q-ansatz}
\eqalign{
q(t) &= A_0\cos(\omega_c t) + B_0\sin(\omega_c t)\\
&+ A_-\cos((\omega_c - \Omega_m)t) + B_-\sin((\omega_c - \Omega_m)t)\\
&+ A_+\cos((\omega_c + \Omega_m)t) + B_+\sin((\omega_c + \Omega_m)t)}.
\end{equation}

An exact solution for \eqref{eq:charge} can be written down, however, the resulting expression is rather unwieldy and we find the approximate solution more illustrative.

Substituting \eqref{eq:q-ansatz} to \eqref{eq:charge}, and neglecting second-order sidebands at frequencies
$\omega_c\pm 2\Omega_m$ (which are proportional to $\delta C^2$), we construct 6 linear equations for the coefficients $A, B$ of \eqref{eq:q-ansatz} by balancing the terms in front of the sines and cosines. The resulting linear system was solved with a symbolic mathematics package \texttt{SymPy} yielding a solution $q(t)$ and the current $I(t)$. 

The lock-in outputs are then given by $X = \gamma_L\{I(t)\cos(\omega_c t + \phi_0)\}$, and $Y = \gamma_L\{I(t)\cos(\omega_c t + \phi_0)\}$. Substituting \eref{eq:Cy-coupling} and retaining only the terms that oscillate at the mechanical frequency $\Omega_m$, we get to the leading order in $\omega_c C_0 R$, the lock-in outputs are (where we note that the phase $\phi_0$ is not necessarily identical to the lock-in phase $\phi$)
\begin{equation}
\label{eq:X-R}
\fl
\eqalign{
    X_\phi(t)&= \gamma_L\{y(t)\}C_0U_cG_\mathrm{IV}g_C \left[C_0R\left(\frac{1}{2}\Omega_m^2 + \omega_c^2\right)\sin\phi_0 -\frac{1}{2}\omega_c\cos\phi_0 \right]\\
    &+ \gamma_L\{\dot{y}(t)\}C_0U_cG_\mathrm{IV}g_C\left(\frac{3}{2}C_0R\omega_c\cos\phi_0+\frac{1}{2}\sin\phi_0\right) + O(R^2),
    }
\end{equation}
and
\begin{equation}
\label{eq:Y-R}
\fl
\eqalign{
    Y_\phi(t)&= \gamma_L\{y(t)\}C_0U_cG_\mathrm{IV}g_C \left[C_0R\left(\frac{1}{2}\Omega_m^2 + \omega_c^2\right)\cos\phi_0 +\frac{1}{2}\omega_c\sin\phi_0 \right]\\
    &+ \gamma_L\{\dot{y}(t)\}C_0U_cG_\mathrm{IV}g_C\left(-\frac{3}{2}C_0R\omega_c\sin\phi_0+\frac{1}{2}\cos\phi_0\right) + O(R^2),
    }
\end{equation}
from which \eqref{eq:Sxxyy} directly follows.

Additional complication arises from the fact that the mechanical noise driving the mode is also responsible for a part of the background (most likely due to to microphonics on the coaxial cables leading to the device). As a consequence, the background is not simply an additive constant in the power spectral density but interferes with the mechanical motion which results in slight peak asymmetry seen in e.g., Fig.~\ref{fig:feedback}(a). To show this, we consider the response to be of the form
\begin{equation}
    \label{eq:noisy-y}
    y = \chi^{-1} F_\mathrm{noise} + \xi_nF_\mathrm{noise}
\end{equation}
where $\chi^{-1} = \meff(\omega^2 - \Omega_m^2 + i\omega\Gamma)$ is the (feedback-modified) mechanical susceptibility of the Helmholtz mode and $\xi_n = \xi_n' + i\xi_n''$ is the susceptibility of all other background sources not related to the Helmholtz mode and is assumed constant within the bandwidth of the mechanical mode.

The power spectral density that follows from \eref{eq:noisy-y} is
\begin{equation}
    \label{eq:noisy-Syy}
    S_{yy} = \left(|\chi|^{-2} + |\xi_n|^2 + \Re\left(\xi^*\chi^{-1}\right)\right)S_{FF},
\end{equation}
which is the line shape used to fit the peaks in Fig.~\ref{fig:rms-displacement} and calculate the RMS displacement in \eref{eq:rms-y}.

\bibliographystyle{iopart-num}
\bibliography{helmholtz_feedback}

\end{document}